%% file: main.tex
\documentclass[%
 reprint,
 superscriptaddress,
 longbibliography,
 amsmath,amssymb,
 prx
]{revtex4-2}

\usepackage{xcolor}

\usepackage[normalem]{ulem}
\usepackage{array}
\usepackage{graphicx}
\usepackage{dcolumn}
\usepackage{bm}

\usepackage{siunitx}
\usepackage{braket}
\usepackage{hyperref}
\DeclareSIUnit{\dBm}{dBm}
\begin{document}

\preprint{APS/123-QED}

\title{Watching a Superconducting Coplanar Waveguide Heat Up with a Single Color Center}

\newcommand{\kitphi}{Physikalisches Institut (PHI), Karlsruhe Institute of Technology (KIT), Kaiserstraße 12, 76131 Karlsruhe, Germany}

\newcommand{\saarb}{Department of Physics, Saarland University, Campus E2 6, 66123 Saarbrücken, Germany}

\newcommand{\leipzig}{Division of Applied Quantum Systems, Felix-Bloch-Institute for Solid State Physics, University of Leipzig, 04103 Leipzig, Germany}

\newcommand{\kassel}{Institute of Nanostructure Technologies and Analytics (INA), Center for Interdisciplinary Nanostructure Science and Technology (CINSaT), University of Kassel, Heinrich-Plett-Straße 40, 34132 Kassel, Germany}

\newcommand{\kitiqmt}{Institute for Quantum Materials and Technologies (IQMT), Karlsruhe Institute of Technology (KIT), Hermann-von-Helmholtz Platz 1, 76344 Eggenstein-Leopoldshafen, Germany}

\author{Ioannis Karapatzakis}
\email{ioannis.karapatzakis@kit.edu}
\affiliation{\kitphi}%

\author{Jeremias Resch}
\affiliation{\kitphi}%










\author{David Hunger}
\affiliation{\kitphi}
\affiliation{\kitiqmt}%

\author{Wolfgang Wernsdorfer}
\affiliation{\kitphi}
\affiliation{\kitiqmt}%

\date{\today}

\begin{abstract}
Single color centers in diamond offer a local probe of their cryogenic environment, providing a direct way to quantify heating in spin-control hardware. Here, we establish a single spectrally stable tin-vacancy (SnV) center as an on-chip thermometer for a diamond membrane and use it to characterize microwave- and radio-frequency-induced heating in a superconducting coplanar waveguide patterned on the same chip.
We first calibrate the temperature dependence of the optical C-transition frequency and linewidth from \SI{20}{\kelvin} down to the few-kelvin regime. At lower temperatures, where the optical response becomes weakly temperature dependent, we use the spin--lattice relaxation time $T_1$ as a complementary thermometer and tune its sensitivity with the transverse magnetic-field component.
Applying this local thermometer to a niobium coplanar waveguide, we observe magnetic-field-dependent superconducting breakdown under GHz drive, accompanied by abrupt heating of the diamond.
In contrast, at \SI{20}{\mega\hertz} and \SI{400}{\milli\tesla}, relevant for nuclear-spin control, we detect no measurable heating up to the breakdown threshold of \SI{9.4}{\dBm}, corresponding to $B_\text{ac} \sim \SI{1.2}{\milli\tesla}$. 
These results define a safe operating window for superconducting microwave and RF control structures in diamond-based quantum nodes.
\end{abstract}

\maketitle


\section{\label{sec:introduction}Introduction}

Optically addressable spin qubits in diamond are promising building blocks for quantum networks~\cite{Bernien_2013, Kalb_2017, Pompili_2021, Stas_2022, Knaut_2024} and quantum information processing, but their performance is critically linked to the local cryogenic environment~\cite{Harris_2024}, where even modest thermal excitation can deteriorate spin initialization fidelity, gate fidelities~\cite{Gundlapalli_2025, Grimm_2025}, and coherence times~\cite{Nguyen_2019}.
Efficient and coherent microwave (MW) and radio-frequency (RF) control of such solid-state spin qubits requires careful engineering of the MW/RF delivery structures, especially in a cryogenic environment, because MW and RF dissipation in normal-conducting spin-control elements such as transmission lines, on-chip waveguides~\cite{Guo_2023}, or bonding wires~\cite{Beukers_2025, Rosenthal_2023} is typically the dominant source of heat load.
This is particularly problematic in diamond, whose extremely low heat capacity at sub-Kelvin temperatures, arising from its very
high Debye temperature and $T^3$ phonon heat capacity~\cite{Vasiliev_2010, DeSorbo_1953, Reeber_1996}, makes the system highly susceptible to even small amounts of dissipation, posing a severe challenge for the fabrication and operation of on-chip normal-conducting transmission-line structures.

In earlier work, we mitigated this by using a superconducting coplanar waveguide (SC-CPW) to control the electron spin of a tin-vacancy (SnV) center in diamond~\cite{karapatz2024} and a proximal $^{13}$C nuclear spin~\cite{resch2025}, strongly reducing MW-induced heating.
To quantify the remaining MW- and RF-induced heating in such devices, a thermometer is required that probes the temperature locally, in the same diamond sample and under the same magnetic-field and drive conditions used for spin control.
Group-IV defects in diamond are well suited for this purpose with narrow and bright~\cite{Neu_2011, trusheim2020transform}, temperature-dependent zero-phonon lines (ZPLs)~\cite{Razgulov2021, Razgulov2022, Neu2013, Arend, Sedov, FanGeV, Goerlitz_2020, Wang_pbv}. So far, group-IV-based thermometry has predominantly relied on the temperature dependence of ensemble photoluminescence (PL) studies of the ZPL, an approach that is limited by inhomogeneous broadening of the ensemble and finite spectral resolution, restricting the accuracy at low temperatures. What is missing is a calibrated, local, single-defect thermometer that works in the same device and field configuration and remains usable from the few-kelvin regime down to the millikelvin regime.

In this work, we establish a single tin-vacancy (SnV) center as such a local thermometer. We use photoluminescence excitation (PLE) spectroscopy of the optical C-transition to calibrate the temperature-dependent lineshift and linewidth from \SI{20}{\kelvin} to the few-kelvin regime. Below this range, where the optical response becomes less sensitive, we use the temperature dependence of the spin--lattice relaxation time $T_1$ of the well-controlled electron spin-qubit in an off-axis magnetic field. This combined optical and spin-relaxation thermometry allows us to probe heating in the same device geometry used for microwave and RF control.

We apply this SnV-based thermometer to quantify MW- and RF-induced heating from a niobium SC-CPW patterned on the same diamond, identifying residual Ohmic losses and the onset of superconductivity breakdown under drive.
We find that the SC-CPW can be operated in a strong magnetic field of \SI{400}{\milli\tesla} up to input powers in the \SI{10}{\dBm} range for both MW and RF frequencies, which for our device geometry corresponds to $B_\text{ac} \sim \SI{1.2}{\milli\tesla}$ at the SnV center.
This demonstrates the suitability of niobium SC-CPWs for MW-(RF-)based electron (nuclear) spin manipulation.

\begin{figure*}
\centering
\includegraphics{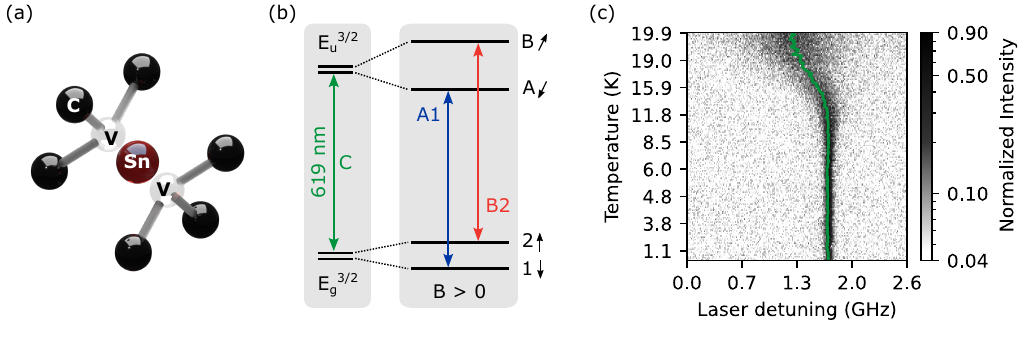}
\caption{\label{fig:Figure_1}
(a) Rendered image of the SnV center in diamond.
(b) Reduced energy level diagram of the SnV center. In a magnetic field, spin degeneracy is lifted. The corresponding transitions are labeled as A1 (blue) and B2 (red).
(c) PLE measurement of the C-transition of SnV-F as a function of temperature. The 2D plot shows the spectral response from $\SI{20}{\kelvin}$ down to $\SI{1}{\kelvin}$, illustrating the evolution of the transition linewidth and position. In order to better visualize the transitions at elevated temperatures, the data have been normalized globally and plotted on a logarithmic scale.
The center position of the transition is indicated with a green line. The frequency axis is offset by \SI{484.170883}{\tera\hertz}. A constant excitation power of \SI{2}{\nano\watt} was used for this measurement.
}
\end{figure*}

\section{\label{sec:setup}Temperature calibration of the SnV center}
In the following, we focus on a single negatively charged tin-vacancy (SnV$^-$) center forming an effective spin-1/2 system. For the experiment reported here, the defect is created by implantation of $^{116}\text{Sn}$ followed by high-temperature annealing.
It consists of a split-vacancy configuration with $D_{3d}$ symmetry, in which the Sn atom is located between two vacant carbon sites, see Fig.~\ref{fig:Figure_1}(a).
The relevant electronic structure can be reduced to the lower ground- and excited-state doublets, $E_\text{g}^{3/2}$ and $E_\text{u}^{3/2}$, separated by \SI{619}{\nano\meter} (green arrow), as sketched in Fig.~\ref{fig:Figure_1}(b).

At cryogenic temperatures, we use the optical C-transition for PLE-based thermometry.
At zero magnetic field this transition is spin degenerate and appears as a single optical resonance.
When a magnetic field is applied, the spin degeneracy is lifted and the C-transition separates into two spin-conserving optical transitions, labeled A1 (blue) and B2 (red) following the notation used in Refs.~\cite{karapatz2024,resch2025}.
These two transitions provide both the optical access needed for spin readout and the optical pumping mechanism used below to measure the electron's spin--lattice relaxation time $T_1$.

The temperature-dependent shift and broadening of the C-transition of the low-strain SnV center SnV-F are shown in the PLE measurement in Fig.~\ref{fig:Figure_1}(c).
The data were recorded while slowly cooling from 20 down to \SI{1}{\kelvin}. The temperature axis is nonlinear and reflects the actual temperature-versus-time curve during the cooldown.

The spectral response is normalized globally and displayed on a logarithmic scale with an adapted range of the colormap, to illustrate the nuanced spectral features at higher temperature.
As the temperature decreases, the C-transition shifts to blue frequencies.
The temperature dependency of the ZPL shift can be attributed to the modulation of the electronic states due to enhanced electron-phonon interaction, but in parts also to the thermal expansion of the diamond~\cite{Razgulov2021}.
The center of the ZPL transition is marked with a green line, as a guide to the eye.

\begin{figure}
\centering
\includegraphics{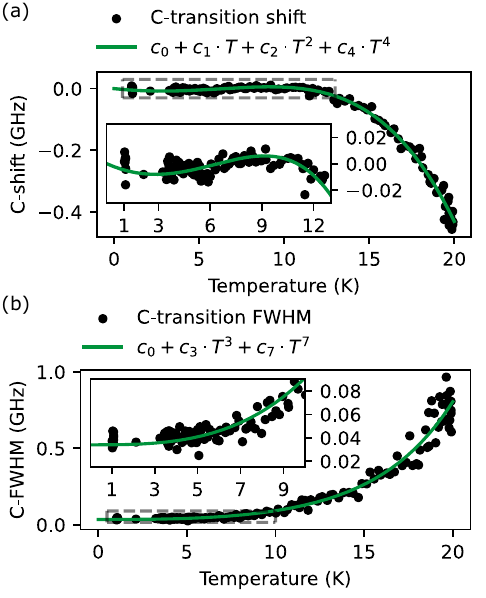}
\caption{\label{fig:Figure_2}
Temperature dependence of the C-transition lineshift (a) and linewidth (b). The black dots show the extracted lineshift/linewidth from the PLE measurements in Fig~\ref{fig:Figure_1}(c).
In (a), the main figure shows a pronounced lineshift above $\SI{12}{\kelvin}$.
The solid green lines show the polynomial fit ($c_0 + c_1 T +  c_2 T^2 + c_4 T^4$).
The inset zooms in on the low-temperature regime from $1-\SI{13}{\kelvin}$, highlighting the positive $T^2$ contribution.
In (b), the main figure reveals a broadening above $\SI{5}{\kelvin}$.
The solid green line shows the fit using a polynomial model $c + c_3 T^3 + c_7 T^7$.
The inset zooms in on the low-temperature region from $1-\SI{10}{\kelvin}$, highlighting the low-temperature precision of the polynomial fits.
}
\end{figure}

Fig.~\ref{fig:Figure_2}(a) and (b) show the extracted ZPL shift and linewidth, plotted against temperature and fitted to a polynomial model of the form $c_0 +c_1T + c_2T^2 + c_4T^4$ and $c + c_3T^3 + c_7T^7$, respectively.
A discussion about the origin of the polynomials and a comparison to other studies of group-IV centers is given in Appendix~\ref{appsec:origin_poly} and Table~\ref{apptab:Temp_group4}.
The tabulated fit values of the lineshift and linewidth are found in Table~\ref{apptab:ZPL_fit_params}.

The inset of Fig.~\ref{fig:Figure_2}(a) shows that the lineshift of the SnV center remains largely insensitive below $\sim\SI{12}{\kelvin}$. In this regime, the positive $T^2$ term is partially compensated by a negative $T^4$ contribution.
In contrast, the extracted linewidth [Fig.~\ref{fig:Figure_2}(b)] increases monotonically with temperature.
The inset highlights the low-temperature region below \SI{10}{\kelvin}, showing that the linewidth already begins to broaden above about \SI{5}{\kelvin}.
This steady and predictable broadening makes the linewidth a more dependable parameter for temperature estimation in this temperature regime.
Thus, the two optical observables provide complementary thermometers. The lineshift is most useful above approximately $\SI{12}{\kelvin}$, whereas the linewidth remains monotonic and more sensitive in the lower few-kelvin regime.
However, below this range, both optical observables become progressively less sensitive.

\begin{figure}
\centering
\includegraphics{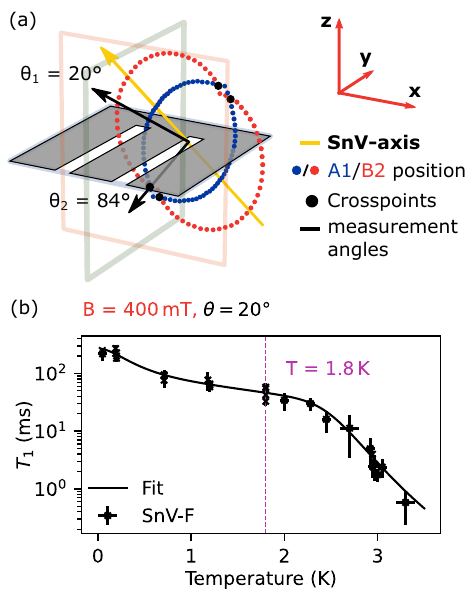}
\caption{\label{fig:Figure_3}
(a) Schematic illustration of the allowed transitions A1 (blue) and B2 (red) of SnV-F extracted from angular sweeps and the relative orientation of the transitions with respect to the SC-CPW structure (grey CPW) on the diamond. The SnV axis is indicated by the yellow arrow, while the black arrows denote the external magnetic field orientations applied in subsequent measurements. For $\theta_2 = \SI{84}{\degree}$, the allowed transitions intersect (black dots). The lab frame is represented by the red coordinate system.
(b) 
Temperature dependence of the spin-lattice relaxation time $T_1$ for a magnetic field of \SI{400}{\milli\tesla} at an angle of $\theta_1 = \SI{20}{\degree}$.
Around $\SI{2.5}{\kelvin}$, $T_1$ exhibits saturation-like behavior due to coupling to resonant single-phonon transitions.
}
\end{figure}

To bridge the gap down to the millikelvin regime, we use the temperature dependence of the electron's spin–lattice relaxation time $T_1$ as an additional metric. The magnetic-field orientation provides an important control mechanism. A transverse field component mixes the spin eigenstates and thereby enhances phonon-mediated spin relaxation (see Appendix~\ref{app:SnV-F_char}). By choosing an off-axis configuration, we intentionally shorten $T_1$ into an experimentally accessible range while retaining a strong temperature dependence.
Additionally, the angular dependence of the optical and qubit resonance frequencies with respect to the magnetic-field orientation allows us to determine both the strain magnitude of the SnV center and the orientation of its quantization axis within the diamond lattice. For a detailed discussion of these angular sweeps and the full Hamiltonian characterization, we refer to our earlier work~\cite{karapatz2024}.

Fig.~\ref{fig:Figure_3}(a) schematically projects the used SnV center into the CPW geometry, showing the SnV-F axis (yellow), the niobium SC-CPW (grey), and the laboratory frame defined by the three-axis vector magnet (red coordinate system).
The niobium SC-CPW is fabricated on the diamond by optical lithography and deposition of a \SI{50}{\nano\meter}-thick Nb film using electron-beam evaporation. The waveguide features a \SI{5}{\micro\meter} gap at the constriction and \SI{30}{\micro\meter}-wide center and ground planes.
The blue (A1) and red (B2) dots indicate the frequency offsets of the two transitions as a magnetic field of fixed magnitude $B = \SI{200}{\milli\tesla}$ is rotated around the defect.
The black arrows mark the two DC magnetic-field orientations used in the subsequent $T_1$ measurements.

Fig.~\ref{fig:Figure_3}(b) shows the temperature dependence of $T_1$ measured at a magnetic field of $B = \SI{400}{\milli\tesla}$ and an angle of $\theta_1 = \SI{20}{\degree}$.
This off-axis configuration suppresses $T_1$ by approximately two orders of magnitude compared to a field parallel to the SnV axis, reducing $T_1$ to values that can be measured within a practical acquisition time and allowing an accurate characterization of the spin–lattice relaxation.
For example, at \SI{1.8}{\kelvin} (indicated by the purple vertical line), we find $T_1 \approx \SI{45}{\milli\second}$, and we use this operating point as the base temperature for the $T_1$ measurements in the following section on the SC-CPW characterization.

At around \SI{2.5}{\kelvin}, $T_1$ shows a slower increase due to coupling to resonant single-phonon transitions.
At temperatures below approximately \SI{200}{\milli\kelvin}, the spin relaxation time is expected to increase substantially because the population of thermally occupied phonon modes decreases polynomially with temperature.
However, we observe a saturation of $T_1 \approx\SI{270}{\milli\second}$, indicating a higher sample temperature.
We emphasize that this apparent low-temperature saturation reflects the local thermal environment of the present setup rather than an intrinsic limit of the SnV center thermometer.
In particular, heat radiation from the \SI{4}{\kelvin} environment, including the objective and the magnetic-field coils mounted on the radiation shields, effectively limits the minimum achievable local sample temperature.
Improved radiation shielding and reduced optical access should offer operation at substantially lower local temperatures, with transverse magnetic-field tuning enabling $T_1$-based sensitivity into the low-millikelvin regime.

To estimate the temperature offset of the sample to the measured temperature $T_\text{m}$ of the thermometer placed at the millikelvin-plate, we introduce a correction term $\Delta T$ and define an effective temperature $T_\text{eff}$ over an exponentially decaying offset term:
\begin{equation}
    T_\text{eff} = T_\text{m} + \Delta T \cdot e^{-\frac{T_\text{m}}{\Delta T}}
\end{equation}
This term accounts for the decreasing influence of thermal radiation as the cooling power increases at higher temperatures and serves as a quantitative measure.
We fit $T_1$ using the spin relaxation model described by S. Meesala et al.~\cite{Meesala_strain}. The tabulated fit values are summarized in Table~\ref{apptab:T1-fits}.
From the best fit, we obtain a sample base temperature of $\SI{336\pm22}{\milli\kelvin}$.
These calibrated optical shifts and relaxation times now enable us to measure local heating caused by microwave delivery through the SC-CPW.

\section{\label{MW_losses}Microwave-Induced Heating in SC-CPWs}

\begin{figure*}
\centering
\includegraphics{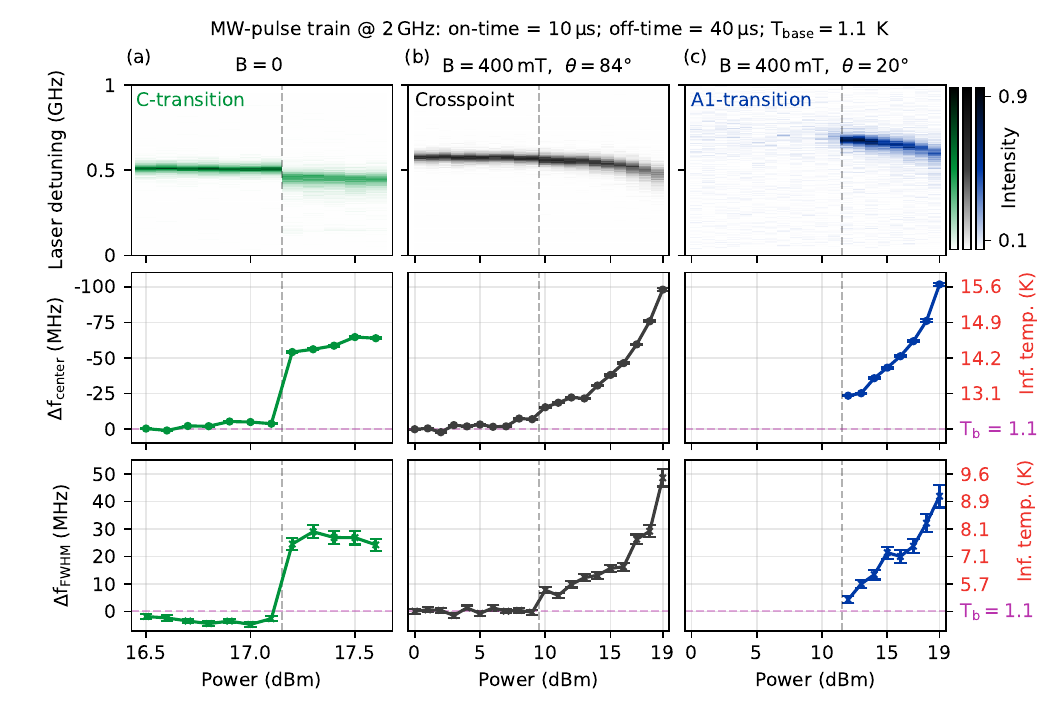}
\caption{\label{fig:Figure_4}
Two-dimensional PLE maps (top row) recorded as a function of laser detuning and applied input power for a MW-pulse train at $f_\mathrm{bias}=\SI{2}{GHz}$ (on-time \SI{10}{\micro\second}, off-time \SI{40}{\micro\second}) and a base temperature $T_\mathrm{base}=\SI{1.1}{K}$.
(a) $B=0$: the degenerate C-transition remains visible throughout the scan and is tracked continuously.
(b) $B=\SI{400}{mT}$, $\theta=\SI{84}{\degree}$: scan around the A1/B2-crosspoint.
(c) $B=\SI{400}{mT}$, $\theta=\SI{20}{\degree}$: lifted spin degeneracy.
Below the breakdown threshold the A1-transition is suppressed by optical spin pumping and becomes visible once heating shortens $T_1$.
Grey dashed lines indicate the critical power $P_\mathrm{crit}$ at which the superconducting state collapses.
Middle row: extracted center-frequency shift $\Delta f_\mathrm{center}$.
Bottom row: extracted linewidth change $\Delta f_\mathrm{FWHM}$.
Right axes convert the measured spectral changes into an inferred local temperature using the calibration in the previous section.
The magenta dashed line marks $T_\mathrm{base}$.
}
\end{figure*}

Having established the SnV center as a calibrated on-chip thermometer, we now use it to quantify microwave- and radio-frequency–induced heating in a SC-CPW patterned on the same diamond.
In typical spin-qubit experiments, dissipation in the MW/RF-delivery structures is a critical factor for device performance, especially under high-power operation.
Here, we use SnV-F as a local sensor to monitor the diamond temperature and to observe MW/RF-related heating and breakdown of superconductivity in the SC-CPW by tracking shifts and broadening of the C- and A1-transitions in PLE, and, where necessary, by measuring the spin–lattice relaxation time $T_1$ of the qubit.

The thermal response of the system is governed by a combination of the heat capacity, the thermal conductivity and finally by the thermal diffusivity.
At cryogenic temperatures the volumetric heat capacity of diamond follows Debye's law, $C_v \propto T^3$, and the thermal conductivity is given by
$\kappa \simeq \tfrac{1}{3} C_v \bar v \ell$, where $\bar v \sim\SI{13}{\kilo\meter\per\second}$ is the average sound velocity and $\ell$ is the effective phonon mean free path~\cite{McSkimin_Bond_1957}.
For electronic grade diamond with low concentration of impurities, phonon scattering is boundary-limited (Casimir regime) and $\ell$ is set mainly by the membrane geometry of $\SI{24}{\micro\meter}\times\SI{1}{\milli\meter}\times\SI{1}{\milli\meter}$~\cite{Efimov_1999}.
Under these conditions, heat transport within the diamond membrane is dominated by thermal diffusivity
$\alpha=\kappa/C_v \simeq \tfrac{1}{3}\bar v \ell$ and is approximately temperature-independent at cryogenic temperatures.
Various groups have investigated the heat capacity~\cite{DeSorbo_1953, Desnoyehs_1958, Vasiliev_2010, Reeber_1996} and conductivity~\cite{Inyushkin_2018, Morelli_1988,Anthony_1991} of diamond. For single crystal diamond the resulting thermal diffusivity can exceed $\SI{5}{\square\meter\per\second}$ below \SI{10}{\kelvin}, matching well with theory.
As heat is generated over the entire area of the SC-CPW structure on the diamond surface, it spreads quickly throughout the membrane. The characteristic timescale for in-plane heat diffusion in the Kelvin regime (i.e., temperatures of a few kelvin) can be estimated by solving the heat equation. Taking the in-plane length scale as $L=\SI{1}{mm}$, the lateral equilibration time is
$\tau_{\mathrm{diff}}\sim L^2/\alpha \approx \SI{0.2}{\micro\second}$. Accordingly, equilibration across the diamond thickness of $t=\SI{24}{\micro\meter}$ will be orders of magnitude faster.

However, while the diamond membrane equilibrates internally on sub-\si{\micro\second} timescales, the overall thermalization of the sample is instead governed by the thermal link to the copper sample holder via aluminum bonding wires and, more importantly, by the thermal mass of the UV-cured adhesive layer of thickness $d_{\mathrm{epoxy}}\simeq\SI{100}{\micro m}$ used to bond the membrane.
Since the exact cryogenic thermal properties of this urethane-related resin (NOA 63~\cite{noa63}) are not available, we estimate them using epoxy and cyanate-ester resins as proxies~\cite{Kelham_Rosenberg_1981, Nakamura_2018}.
On that basis, the adhesive's specific heat capacity at cryogenic temperatures is approximately three orders of magnitude larger than that of diamond. Combined with its extremely low thermal diffusivity, $\alpha_{\mathrm{epoxy}}\sim 10^{-6}$--$10^{-3}\,\mathrm{m^2 s^{-1}}$~\cite{Voorde_1976, Kelham_Rosenberg_1981}, this leads to relaxation times on the order of milliseconds.

The measurements shown in Fig.~\ref{fig:Figure_4} are performed at a base temperature of $\SI{1.1}{\kelvin}$. 
We apply a repetitive microwave pulse at a frequency of $\SI{2}{\giga\hertz}$ with variable power while tracking optical transitions via PLE under various measurement conditions.
The MW-pulse train is detuned from all relevant spin transitions, while remaining within the typical electron spin resonance (ESR) frequency range.
The pulse train has a $\SI{20}{\percent}$ duty cycle, consisting of a $\SI{10}{\micro\second}$ on-time followed by a $\SI{40}{\micro\second}$ off-time, corresponding to a repetition rate of $f=\SI{20}{\kilo\hertz}$. The pulse on-time is chosen to be much shorter than the measured (epoxy-limited) thermalization time of the diamond membrane in the order of \SI{0.5}{\milli\second} (see Appendix~\ref{app:thermalization} and Fig.~\ref{appfig:Figure_7}).
As a result, the membrane temperature follows the time-averaged dissipated microwave power.
The system reaches a steady state within a few cycles and the membrane temperature can be treated as approximately constant, with only a small residual temperature modulation at the repetition frequency. Consistent with the fast internal heat diffusion discussed above, spatial temperature gradients within the membrane are likewise expected to be negligible.

At first, the C-transition under zero magnetic field and its linewidth are tracked, as this transition remains visible during PLE measurements at all times, independent of temperature, due to its degeneracy. 
Fig.~\ref{fig:Figure_4}~(a) presents the corresponding PLE map, where the horizontal axis shows the applied MW input power (referenced at the cryostat input) and the vertical axis denotes the laser detuning. 
Up to $P_\mathrm{crit}=\SI{17.2}{\dBm}$ (grey dashed line), the transition remains Fourier-limited.
Above this threshold, the resonance shifts to lower frequencies and the linewidth broadens.
We attribute this abrupt change to a breakdown of superconductivity in the niobium CPW.
The increased MW drive raises the local current density in the conductor, and when the equivalent peak current approaches the critical current for the relevant cross section, superconductivity is locally quenched.
The niobium waveguide becomes normal conducting, leading to a significant increase in Ohmic MW losses and consequent heating of the diamond.
For better comparison, the extracted center-frequency shift $\Delta f_\mathrm{center}$ and linewidth change $\Delta f_\mathrm{FWHM}$ are shown in the second and third rows, respectively.
Using the calibration of the C-transition established in the previous section, we convert both observables into an effective local temperature with the corresponding temperature scales indicated on the far right-hand side of the axes.
From these conversions we infer an abrupt temperature rise from the base temperature of \SI{1.1}{\kelvin} to $\sim\SI{14}{\kelvin}$ based on the lineshift and to $\sim\SI{7.5}{\kelvin}$ based on the linewidth broadening once superconductivity is lost.
While the linewidth is expected to mainly reflect phonon-induced broadening, the different inferred temperatures indicate that, in the pulsed MW regime after superconducting breakdown, the center frequency is additionally governed by the (alternating) local strain environment~\cite{Brevoord_2026}.
Because the overall thermal relaxation is limited by the adhesive layer on a millisecond timescale, the diamond membrane can heat during each breakdown pulse faster than the adhesive can equilibrate.
The resulting temperature mismatch between the diamond and the mechanically coupled adhesive layer can generate thermoelastic strain in the membrane.
The temperature obtained from the lineshift can therefore differ from the temperature inferred by the linewidth, with the latter providing an estimate of the local phonon bath temperature.

Next, we repeat the measurement in a magnetic field of $B=\SI{400}{\milli\tesla}$ applied at an angle of \SI{84}{\degree} and \SI{20}{\degree} with respect to the SnV-F quantization axis, shown in columns (b) and (c) of Fig.~\ref{fig:Figure_4}, respectively. Please note the increased MW input-power range of $P_\mathrm{in}=\SIrange{0}{19}{\dBm}$.
For the configuration in (b), the A1- and B2-transitions cross (see Fig.~\ref{fig:Figure_3}~(a)), and the overall behavior remains analogous to the zero-field case.
However, the superconducting breakdown occurs already at a reduced threshold power of $P_\mathrm{crit}\sim\SI{10}{\dBm}$, as the applied magnetic field partially suppresses superconductivity in the niobium film~\cite{janjuvsevic2006microwave}.
The extracted $\Delta f_\mathrm{center}$ and $\Delta f_\mathrm{FWHM}$ in rows two and three exhibit the same qualitative behavior as in panel~(a), as both remain constant below $P_\mathrm{crit}$ and increase once superconductivity breaks down.

In configuration~(c), where the spin degeneracy is lifted, the laser is scanned across the A1-transition and optically pumps population from $\ket{\downarrow}$ into $\ket{\uparrow}$.
As a consequence, the system is prepared in the state addressed by B2, and the A1-transition becomes dark.
As long as superconductivity is maintained and the local temperature remains sufficiently low ($<\SI{4}{K}$), such that $T_1$-mediated spin relaxation is negligible on the measurement timescale, no PLE signal associated with A1 is observed.
In this configuration, superconductivity breaks down at an input power of $P_\text{crit}\sim\SI{12}{\dBm}$, at which the A1-transition becomes immediately visible, along with a pronounced red shift and linewidth broadening.
Although the same field magnitude ($B=\SI{400}{\milli\tesla}$) is applied as in panel~(b), the breakdown occurs at a higher MW power here.
The observed behavior is consistent with vortex-induced losses in the type-II niobium film.
In the presence of a strong magnetic field, magnetic flux penetrates the SC-CPW in the form of Abrikosov vortices~\cite{buckel2013supraleitung}, and the superconductor enters the mixed state~\cite{matsushita2007flux}.
The normal-conducting vortex cores can be driven by the Lorentz force from the oscillating transport current, and this vortex motion gives rise to flux-flow dissipation~\cite{bardeen1965theory, kim1965flux, janjuvsevic2006microwave}.
Material defects and inhomogeneities pin the vortices and suppress their motion for sufficiently small ac fields and for drive frequencies below a characteristic pinning frequency~\cite{park1992vortex}.
The number of vortices depends on the perpendicular field component, which is smaller for configuration (c) i.e., the field is more in-plane relative to the superconducting film~\cite{janjuvsevic2006microwave}.

\begin{figure*}
\centering
\includegraphics{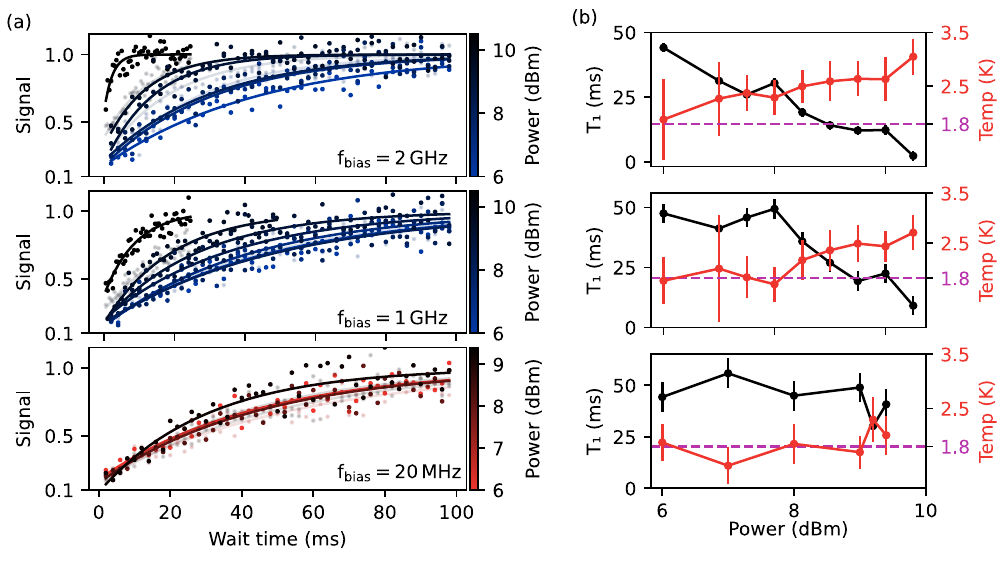}
\caption{\label{fig:Figure_5}
Measurements performed in the field configuration of Fig.~4(c) ($B=\SI{400}{mT}$, $\theta=\SI{20}{\degree}$) at $T_\mathrm{base}=\SI{1.8}{K}$ while sweeping the applied input power.
The bias is applied with \SI{100}{\percent} duty cycle during the $T_1$ sequence.
(a) Representative spin-relaxation traces for three bias frequencies ($f_\mathrm{bias}=\SI{2}{\giga\hertz}$, \SI{1}{\giga\hertz}, and \SI{20}{\mega\hertz}) and different input powers (color-coded).
The solid lines are fits used to extract $T_1$.
(b) Extracted $T_1$ versus input power (black, left axis) together with the corresponding inferred local temperature obtained from the $T_1(T)$ calibration (red, right axis).
The magenta dashed line indicates $T_\mathrm{base}$.
Data terminate once superconductivity collapses and $T_1$ measurements are no longer attainable.
}
\end{figure*}

To gain further insight into the behavior close to the critical power threshold, we measure $T_1$ under configuration~(c) at a base temperature of \SI{1.8}{\kelvin} while sweeping the applied MW/RF-input power over \SIrange{6}{11}{\dBm}.
In contrast to the PLE measurements, we use a duty cycle of \SI{100}{\percent} (continuous-wave drive during the $T_1$ sequence), such that any additional heating or suppression of superconductivity is maximized.
At \SI{1.8}{\kelvin}, the relaxation time is reduced to $T_1=\SI{46.9\pm3}{\milli\second}$ [Fig.~\ref{fig:Figure_3}(b)], which allows each data point to be acquired on a timescale of tens of minutes. 
The corresponding $T_1$ traces for three bias frequencies ($\SI{2}{\giga\hertz}$, $\SI{1}{\giga\hertz}$, and $\SI{20}{\mega\hertz}$) are shown in Fig.~\ref{fig:Figure_5}(a).
From these data we extract $T_1$ and convert it into an effective local temperature using the previously established $T_1(T)$ calibration shown in Fig.~\ref{fig:Figure_3}(b). Fig.~\ref{fig:Figure_5}(b) summarizes the result in a dual-axis representation. The measured $T_1$ values are plotted on the left axis (black), while the
corresponding effective sample temperature is given on the right axis (red), both as a function of the applied input power.

For a \SI{2}{\giga\hertz} bias, $T_1$ starts to decrease above $\sim\SI{7}{\dBm}$, indicating a gradual increase of the local temperature from \SI{1.8}{\kelvin} to approximately \SI{3}{\kelvin} as the power is raised to \SI{11}{\dBm}.
At higher powers, $T_1$ measurements are no longer attainable because superconductivity collapses.
The reduced breakdown temperature compared to bulk niobium with $T_\text{c} = \SI{9.25}{\kelvin}$ is expected, as the applied field of \SI{400}{\milli\tesla} reduces the effective critical temperature of the niobium film~\cite{finnemore1966superconducting, hudson1971superconducting} and the additional microwave current further suppresses superconductivity, so that the combination of magnetic field, MW current, and heating leads to a collapse of the superconducting state once the local temperature reaches $\sim\SI{3}{\kelvin}$.
A qualitatively similar behavior is observed for a \SI{1}{\giga\hertz} bias (second row).

We finally apply an RF bias of \SI{20}{\mega\hertz}, which is particularly relevant for nuclear-spin control requiring resonant RF driving.
The corresponding data are shown in the third row of Fig.~\ref{fig:Figure_5}(b).
Remarkably, $T_1$ remains constant within the experimental uncertainty over the full accessible power range, and we do not observe a gradual degradation indicative of progressive heating.
Instead, superconductivity breaks down abruptly at $P_\text{max}^\text{RF}=\SI{9.4}{dBm}$ within a power increment of \SI{0.1}{\dB}, after which $T_1$
measurements are no longer attainable.
This distinct behavior is consistent with operation in a regime where vortex motion is strongly pinned~\cite{matsushita2007flux} and dissipation remains negligible up to a sharp threshold.
With increasing drive, the combination of vortex dynamics, the transport current density, and the field-induced reduction of the effective critical current eventually pushes the film into the normal state.
The lower breakdown power at \SI{20}{\mega\hertz} compared to the microwave case can be understood from the frequency-dependent insertion loss of the coaxial lines.
Since the CPW is short-terminated, we characterize the coupling in reflection via $S_{11}$.
In the lossless limit one expects $|S_{11}|=1$ (input return loss (IRL) $=\SI{0}{\dB}$), while deviations from unity quantify power that is not returned and is instead dissipated mostly along the coaxial line to the sample.
At \SI{2}{\GHz} we measure an IRL of $\sim\SI{2.8}{\dB}$, whereas at \SI{20}{\mega\hertz} the IRL is only $\sim\SI{0.5}{\dB}$, implying that
nearly all of the applied power is delivered to the SC-CPW at \SI{20}{\mega\hertz}. As a result, the on-chip current amplitude reaches the breakdown condition at a smaller quoted input power in the RF case.

Using the independent calibration of the RF field at the SnV position~\cite{resch2025}, the threshold $P_\text{max}^\text{RF}=\SI{9.4}{dBm}$ corresponds to an oscillating magnetic-field amplitude of $B_\mathrm{ac}\approx\SI{1.2}{mT}$. Such a field is sufficient to drive electron-spin Rabi oscillations at \si{\mega\hertz} rates and nearby coupled ${}^{13}$C nuclear spins at \si{\kilo\hertz} rates.
This highlights a key result: the SC-CPW remains fully functional for MW/RF-control even under strong DC-magnetic fields, demonstrating its suitability for electron and nuclear spin manipulation.

\section{\label{sec:conclusion}Conclusion}
In summary, we have shown that a single SnV center can be used as a local probe of the thermal response of a diamond membrane during MW and RF operation of an integrated superconducting CPW.
By combining the temperature-dependent optical response of the C-transition with spin-relaxation thermometry, we cover different temperature regimes with the same defect and in the same device geometry used for spin control.
In particular, the transverse magnetic-field component provides a way to tune $T_1$ into a useful measurement range, extending the thermometer sensitivity from the optical few-kelvin regime down to the millikelvin regime.

In the finite magnetic-field environment required for electron--nuclear spin control, the MW and RF response of the CPW are qualitatively different.
Under \si{\giga\hertz} driving, the SnV center detects the increase of the local temperature as the superconducting limit is approached, followed by a collapse of superconductivity, visible through spectral shifts, linewidth broadening, and gradually reduced $T_1$.
The dependence on the magnetic-field orientation points to vortex-induced dissipation in the type-II niobium film.
In contrast, for RF driving at \SI{20}{\mega\hertz} and \SI{400}{\milli\tesla}, $T_1$ and thus temperature, remains unchanged within the experimental uncertainty up to $P_{\mathrm{RF}}=\SI{9.4}{dBm}$, corresponding to $B_{\mathrm{ac}}\simeq\SI{1.2}{\milli\tesla}$ at the SnV position.

The absence of measurable heating before this sharp RF threshold is particularly important for nuclear-spin control, where the required pulses are much longer than typical electron-spin operations.
More generally, the measurements provide a direct local benchmark for superconducting CPWs in diamond-based quantum devices and identify an operating window in which MW and RF control fields can be applied while preserving the cryogenic environment.
This is especially relevant for quantum-network nodes, where long-lived nuclear memories and repeated control sequences, such as dynamical decoupling, require stable local temperatures over extended operation times.

\begin{acknowledgments}
This work was partly supported by the German Federal Ministry of Research, Technology and Space
(Bundesministerium für Forschung, Technologie und
Raumfahrt, BMFTR) within the projects QR.N (Contract
No. 16KIS2186), QR.X (Contract No. 16KISQ004), SPINNING (Contract No. 13N16211), and the
Karlsruhe School of Optics and Photonics (KSOP).
\end{acknowledgments}

\nocite{*}

\include{app}

\bibliography{apssamp}
\end{document}

%% file: app.tex
\appendix
\section{\label{appsec:origin_poly}Temperature-dependent optical response of group-IV centers}

\begin{table*}
\centering
\caption{\label{apptab:Temp_group4}
Temperature behavior of group-IV defects. Abbreviations: ref. = reference; s.e. = single emitter; ens. = ensemble.}
\begin{tabular}{p{1.4cm} p{1cm} p{2.45cm} p{2.2cm} p{2cm} p{2.2cm}}
\hline
\hline
\textbf{Defect center} & \textbf{ref.} & \textbf{Temperature range (K)} & \textbf{Defect type} & \textbf{Linewidth scaling} & \textbf{Lineshift scaling} \\ \hline
\hline
SiV  & \cite{Jahnke2015phonon_siv} & 4--350   & s.e.        & $< \SI{20}{\kelvin}$: $T^1$;\quad $> \SI{20}{\kelvin}$: $T^3$            & $T^3$ \\[1mm]
SiV  & \cite{Neu2013}             & 5--295   & s.e. and ens.  & $T^3$                           & ens.: $T^2+T^4$; s.e.: $T^4$ \\[1mm]
SiV  & \cite{Arend}              & 5--300   & ens.             & $T^3+T^5+T^7$                         & $T^2+T^4$ \\[1mm]
SiV  & \cite{Sedov}              & 5--80    & ens.             & $T^3+T^5$                             & $T^2+T^4$ \\[1mm]
GeV  & \cite{Razgulov2022}        & 20--180  & ens.             & $T^3$                                 & $T^2+T^4$ \\[1mm]
GeV  & \cite{FanGeV}             & 100--400 & ens.             & $T^{a}$ ($a\approx 2.8$)         & $T^3$ \\[1mm]
SnV  & \cite{Razgulov2021}        & 80--300  & ens.             & $T^3+T^7$                             & $T^2+T^4$ \\[1mm]
SnV  & \cite{Goerlitz_2020}        & 5--200   & s.e. and ens.  & $T^3$                           & $T^2+T^4$ \\[1mm]
PbV  & \cite{Wang_pbv}           & 6--260   & ens.             & $T^3$                                 & $T^2+T^4$ \\ \hline
SnV  & this work   & 1--20   & s.e.  & $T^3+T^7$   & $T+T^2+T^4$ \\
\hline
\hline
\end{tabular}
\end{table*}

The temperature dependence of group-IV color centers, including the SnV center, has been examined in detail by several groups, and an overview of reported results is summarized in Table~\ref{apptab:Temp_group4}. 
In most cases, the ZPL linewidth is found to follow a dominant $T^3$ scaling with an additional weaker $T^7$ contribution, while the ZPL position typically exhibits a mixed $T^2 + T^4$ behavior. 
These empirical power laws can be understood within the electron-phonon coupling model developed by V.~Hizhnyakov~\cite{Hizhnyakov2002,Hizhnyakov2004}, in which long-wavelength acoustic phonons modulate the ZPL frequency. 
For nondegenerate electronic states, this mechanism leads to a $T^7$ scaling of the linewidth, whereas for degenerate states subject to a linear dynamic Jahn-Teller effect, a $T^5$ dependence is expected. 
The latter behavior has so far only been reported in two ensemble studies of the SiV center~\cite{Arend,Sedov}.

Hizhnyakov’s model further incorporates quadratic electron-phonon coupling under conditions of strong softening of interatomic bonds in the excited state. 
For a purely linear Jahn-Teller interaction, the adiabatic potential energy surface (APES) of the excited state has the form of a Mexican-hat potential with a single circular minimum~\cite{Thiering_magneto_optic, magneto_mohseni}. 
When quadratic terms are included, this axial symmetry is lifted and three equivalent minima appear, separated by shallow barriers if the interaction strength is small. 
Such an instability results on a peak of the local phonon density of states at low frequencies, which in turn modifies the temperature scaling of the optical response, leading to a $T^2$ rather than $T^4$ dependence of the ZPL shift and a $T^3$ instead of $T^7$ dependence of the linewidth.

In contrast to most previous studies, our data extend to the \SIrange{1}{4}{\kelvin} regime and require
a small negative linear contribution to reproduce the low-temperature line
shift. We therefore treat $c_1$ as an empirical calibration parameter and
do not assign it to a specific microscopic mechanism.
A possible low-temperature contribution from thermal expansion or residual
stress changes cannot be excluded, but is not required for the thermometer
calibration performed here.

\section{\label{app:SnV-F_char}SnV-F optical and spin thermometry calibration}

The characterization and fit of the electronic energies of SnV-F via the spin-conserving transitions A1 and B2 can be found in Fig.~4 and Table~I of Ref.~\cite{resch2025}.
Here, we focus on the temperature calibration of the C-transition of Figure~\ref{fig:Figure_1}(c). Each PLE trace is fit with a Lorentzian line shape to extract the center frequency and full width at half maximum. The fitted parameters for the C-transition lineshift and linewidth during the cooldown shown in Fig.~\ref{fig:Figure_1}(c) are summarized in Table~\ref{apptab:ZPL_fit_params}.

\begin{table}
\centering
\caption{\label{apptab:ZPL_fit_params}
Fit parameters for the C-transition lineshift and linewidth functions of SnV-F. The uncertainties are given as standard errors.}
\begin{tabular}{c r}
\hline
\hline
\textbf{Parameter} & \textbf{Value}\\
\hline  
\multicolumn{2}{c}{Lineshift: $c_0 + c_1 T +  c_2 T^2 + c_4 T^4$} \\
\hline
$c_0$   & \num{0} (fixed)\\
$c_1$ & \num{-5.87(62)} $\si{\mega\hertz\per\kelvin^1}$\\
$c_2$ & \num{1.11(7)} $\si{\mega\hertz\per\kelvin^2}$\\
$c_4$ & \num{-4.72(12)e-3} $\si{\mega\hertz\per\kelvin^4}$\\
\hline
\hline
\multicolumn{2}{c}{Linewidth: $c + c_3 T^3 + c_7 T^7$} \\
\hline
$c$   & \num{34(4)} $\si{\mega\hertz}$ (see~\cite{resch2025})\\
$c_3$ & \num{54.7(41)e-3} $\si{\mega\hertz\per\kelvin^3}$\\
$c_7$ & \num{26.0(31)e-8} $\si{\mega\hertz\per\kelvin^7}$\\
\hline
\hline
\end{tabular}
\end{table}

The PLE measurements are performed with a resonant laser power of $p = \SI{2}{\nano\watt}$, well below the saturation power of $p_\text{sat} \sim \SI{20}{\nano\watt}$. The data in Fig.~\ref{fig:Figure_2}(a) are centered such that the offset $c_0 = 0$ for the lineshift fit. Temperatures are inferred by numerically inverting the fitted calibration functions for the lineshift and linewidth using the parameters in Table~\ref{apptab:ZPL_fit_params}. Because the lineshift $\Delta f_\mathrm{center}(T)$ is only weakly temperature dependent below approximately $\SI{12}{\kelvin}$, the linewidth provides the more robust optical thermometer in the few-kelvin regime. The extracted zero-temperature linewidth of $c = \SI{35.2 \pm 5.6}{\mega\hertz}$ is slightly larger than the expected Fourier-limited linewidth for SnV centers, $\sim \SI{28}{\mega\hertz}$~\cite{karapatz2024,trusheim2020transform}, which is most likely due to the strong coupling to a nearby $^{13}$C nuclear spin of this SnV center~\cite{resch2025}. Below the temperature range where optical broadening is sufficiently sensitive, we use the spin-lattice relaxation time $T_1$ as an independent thermometer.

\begin{figure}
\centering
\includegraphics{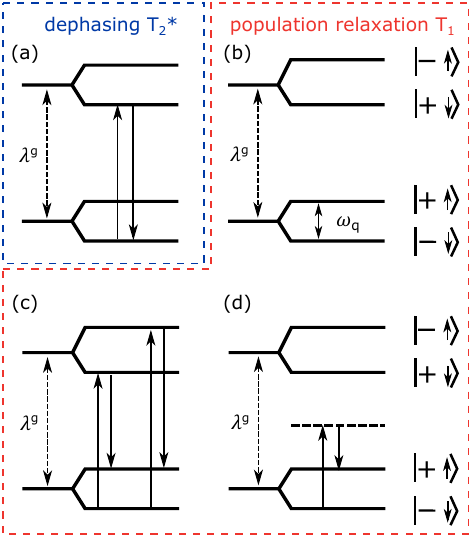}
\caption{\label{appfig:Figure_6}
Phonon-induced relaxation pathways.
The ground-state levels (with splitting $\lambda^\text{g}$) are labeled on the far-right with their respective orbital ($\ket{+,-}$) and spin angular ($\ket{\uparrow,\downarrow}$) momentum.
The blue panel (a) illustrates spin-conserving transitions responsible for pure dephasing. 
The red panel highlights phonon-mediated spin-flip processes: (b) shows direct single-phonon relaxation resonant with the lower qubit frequency $\omega_\text{q}$, (c) depicts two-phonon processes involving the upper orbital branch, and (d) illustrates off-resonant two-phonon relaxation pathways.
}
\end{figure}

Phonon-mediated dephasing in group-IV centers is primarily driven by spin-conserving transitions within the ground-state orbital branches (see Fig.~\ref{appfig:Figure_6}(a), black solid arrows)~\cite{Meesala_strain,Jahnke2015phonon_siv,Harris_2024}. 
The difference in the qubit’s Larmor precession frequency between the two orbital states ($E^\text{g}_{3/2}$ and $E^\text{g}_{1/2}$) leads to an accumulated phase if the qubit is prepared in a superposition state and undergoes phonon-induced excitation for a time $\tau$.

Here, we focus on spin-flip processes, which are primarily governed by the three mechanisms illustrated in Fig.~\ref{appfig:Figure_6}(b--d). 
In Fig.~\ref{appfig:Figure_6}(b), direct single-phonon absorption or emission processes resonant with the qubit frequency $\omega_\text{q}$ lead to electron spin flips. 
In an ideal, unperturbed system, phonon-mediated transitions between orbital eigenstates couple exclusively to the orbital degree of freedom and therefore remain spin-conserving. 
Spin mixing, and thus spin-flip relaxation, occurs only in the presence of an off-axis magnetic field~\cite{Meesala_strain,trusheim2020transform,Debroux2021}. 
Using first-order perturbation theory, the perturbed eigenstates can be expressed as
\begin{align}
|- \downarrow \rangle' &\approx |- \downarrow \rangle - \frac{\gamma_s B_\perp}{2\lambda^\text{g}} |- \uparrow \rangle,\\
|+ \uparrow \rangle' &\approx |+ \uparrow \rangle - \frac{\gamma_s B_\perp}{2\lambda^\text{g}} |+ \downarrow \rangle,
\end{align}
where $\gamma_s = \SI{28}{\giga\hertz\per\tesla}$ is the electron gyromagnetic ratio and $B_\perp$ denotes the transverse component of the applied magnetic field.
Fig.~\ref{appfig:Figure_6}(c) depicts a two-phonon process that mediates a spin flip. 
Analogous to the single-phonon case, a transition such as $\ket{+\downarrow} \rightarrow \ket{+\uparrow}$, which preserves the orbital character, becomes allowed only in a perturbed system with finite local strain and a nonzero off-axis magnetic field.
In Fig.~\ref{appfig:Figure_6}(d), an off-resonant two-phonon process is shown. 
While such relaxation mechanisms are typically expected to become significant only at elevated temperatures (above roughly \SI{40}{\kelvin}), we find that including this contribution is necessary to reproduce the experimental data for SnV-F.

Here, this model is used as a calibrated interpolation between the thermometer temperature and the local SnV temperature.
The prefactors $A_1$, $A_2$, and $A_3$ in Table~\ref{apptab:T1-fits} represent effective rates for the direct, resonant two-phonon, and off-resonant two-phonon contributions under the specific strain and magnetic-field configuration of SnV-F.
The transverse magnetic-field component $B_\perp$ is fixed by the independently determined SnV axis and is the main parameter that tunes the relaxation rate into an experimentally accessible range.

\begin{table}
\centering
\caption{\label{apptab:T1-fits}
Fit values of $T_1$ versus temperature for SnV-F. The parameters $A_1(\text{single-phonon})$, $A_2(\text{two-phonon})$, and $A_3(\text{off-resonant two-phonon})$ represent the amplitudes (pre-factors of the expected scaling rate) for the corresponding transition rates.}
\begin{tabular}{c c c}
    \hline
    \hline
    \textbf{Parameter} & \textbf{SnV-F} \\ \hline
    $A_1$ & $4.04(34) \times 10^{-7}\,\si{\kilo\hertz}$ \\ 
    
    $A_2$ & $1.36(15) \times 10^{-5}\,\si{\kilo\hertz}$ \\ 
    
    $A_3$ & $189(80)\,\si{\kilo\hertz}$ \\ 
    
    $B_{\perp}$ & $\SI{136.8}{\milli\tesla}$ (fixed) \\ 
    
    $\lambda^\text{g}$ & $\SI{840}{\giga\hertz}$ (fixed) \\
    $\omega_\text{q}$ & $\SI{13}{\giga\hertz}$ (fixed) \\
    $\Delta T$ & \SI{336\pm22}{\milli\kelvin} \\
    \hline
    \hline
\end{tabular}
\end{table}

\section{\label{app:device_power_calibration}Device, power conventions, and MW/RF field calibration}

The diamond membrane preparation, Sn implantation, cryogenic confocal setup, and Hamiltonian characterization follow Refs.~\cite{karapatz2024,resch2025}. 
Here we only summarize the details relevant for thermometry and CPW heating.
The niobium CPW has a thickness of \SI{50}{\nano\meter} and a \SI{5}{\micro\meter} gap at the end of the constriction and is patterned directly on the diamond membrane.
The quoted MW and RF powers always refer to the power measured before the cryostat input.
For pulsed measurements, the quoted value is the instantaneous input power during the pulse, while the average dissipated power is reduced by the duty cycle.

Using the independently calibrated RF field amplitude at the SnV center's position from Ref.~\cite{resch2025}, an oscillating magnetic-field amplitude of $B_{\mathrm{ac}}>\SI{1}{\milli\tesla}$ at the defect's location is easily accessible for the used CPW geometry.
This amplitude is large on the scale required for coherent spin control: for an electron spin it corresponds to Rabi frequencies in the MHz range, while for nearby coupled $^{13}$C nuclear spins it enables kHz-scale resonant driving~\cite{resch2025}.
In addition, the different breakdown powers should not be interpreted as intrinsic critical powers of niobium alone.
They depend on the static magnetic-field orientation, the resulting vortex configuration and pinning landscape, the frequency of the applied drive, and the pulse duty cycle.
For SnV centers, the most relevant operating regime for electro-nuclear spin control is at static magnetic fields of order $\sim\SI{100}{\milli\tesla}$, with RF frequencies from near-DC up to $\sim\SI{20}{\mega\hertz}$ for nuclear-spin driving and microwave frequencies in the low-GHz range for electron-spin control. Table~\ref{tab:breakdown_summary} summarizes the MW/RF heating and superconducting-breakdown measurements.

\begin{table*}[tb]
\centering
\caption{\label{tab:breakdown_summary}
Summary of MW/RF heating and superconducting-breakdown measurements.
}
\begin{tabular}{cccccc}
\hline\hline
$f_{\mathrm{bias}}$ & $B$ & angle & duty cycle & observable & threshold/result \\
\hline
\SI{2}{\giga\hertz}  & 0       & --        & \SI{20}{\percent}  & C-transition
       & $P_{\mathrm{crit}}\approx\SI{17.2}{\dBm}$ \\
\SI{2}{\giga\hertz}  & \SI{400}{\milli\tesla}  & $84^\circ$ & \SI{20}{\percent}  & A1/B2 crosspoint
       & $P_{\mathrm{crit}}\approx\SI{10}{\dBm}$ \\
\SI{2}{\giga\hertz}  & \SI{400}{\milli\tesla}  & $20^\circ$ & \SI{20}{\percent}  & A1-transition
       & $P_{\mathrm{crit}}\approx\SI{12}{\dBm}$ \\
\SI{2}{\giga\hertz}  & \SI{400}{\milli\tesla}  & $20^\circ$ & \SI{100}{\percent} & $T_1$
       & gradual heating to $\sim\SI{3}{\kelvin}$ before collapse \\
\SI{1}{\giga\hertz}  & \SI{400}{\milli\tesla}  & $20^\circ$ & \SI{100}{\percent} & $T_1$
       & similar gradual heating trend \\
\SI{20}{\mega\hertz} & \SI{400}{\milli\tesla}  & $20^\circ$ & \SI{100}{\percent} & $T_1$
       & no detectable heating up to $P_{\mathrm{RF}}=\SI{9.4}{\dBm}$ \\
\hline\hline
\end{tabular}
\end{table*}

\section{\label{app:thermalization}Thermal response and validity of the local-temperature extraction}

\begin{figure*}
\centering
\includegraphics{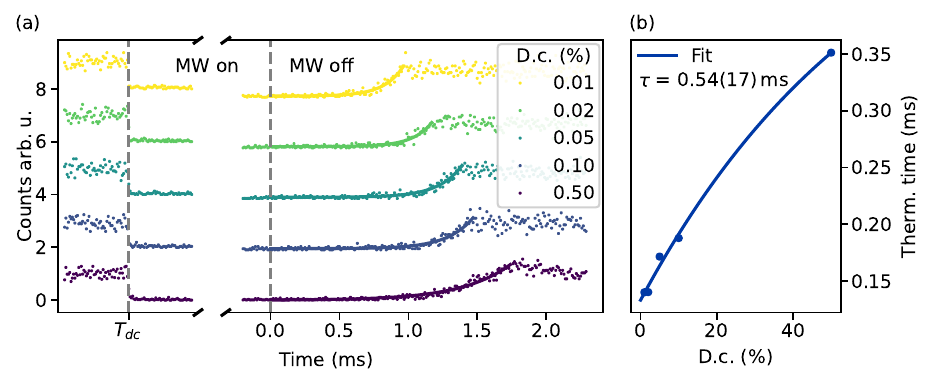}
\caption{\label{appfig:Figure_7}
(a) Time-resolved fluorescence signal recorded while periodically switching the MW bias between an on and off state for different duty cycles (D.c). The traces are vertically offset for clarity.
Grey dashed lines indicate the MW switching times. The recovery during the MW-off segment is fitted with an exponential (solid lines) to extract a characteristic thermal relaxation time.
(b) Extracted thermal relaxation time as a function of duty cycle. The solid line is a fit yielding a quantitative time constant of $\tau=\SI{0.54(17)}{ms}$.
}
\end{figure*}

To determine the relevant thermal timescale of the sample assembly, we periodically switch the MW bias between an on and off state and monitor the time-resolved fluorescence of SnV-F after the MW pulse is turned off.
These measurements, shown in Figure~\ref{appfig:Figure_7}(a), are performed without applied magnetic field and at \SI{1.1}{\kelvin}, such that we probe the degenerate C-transition of the SnV center.
The applied MW power is chosen above the superconducting-breakdown threshold of the CPW.
During the MW-on interval, the niobium line becomes normal conducting and dissipates strongly, heating the diamond membrane. As a result, the C-transition shifts and broadens strongly.
For the fixed laser frequency used in this measurement, the emitter is then no longer resonant and the fluorescence signal is suppressed.
When the MW drive is switched off, the diamond cools back toward the base temperature and the CPW returns to the superconducting state.
The C-transition shifts back into resonance with the laser frequency, producing a recovery of the fluorescence signal.
The fluorescence revival therefore provides a quantitative, local probe of the thermal relaxation of the diamond sample after a heating pulse.

The duty cycle is varied in order to test whether the extracted relaxation time depends on the amount of deposited heat. For duty cycles from $1\%$ to $10\%$, the MW-on time is chosen as $1$--$10\,\si{\milli\second}$ within a fixed $100\,\si{\milli\second}$ period. For example, the $1\%$ trace uses $1\,\si{\milli\second}$ on and $99\,\si{\milli\second}$ off, while the $10\%$ trace uses $10\,\si{\milli\second}$ on and $90\,\si{\milli\second}$ off. The $50\%$ trace is recorded with a shorter total period, using $10\,\si{\milli\second}$ on and $10\,\si{\milli\second}$ off, in order to avoid excessive heating of the cryostat. In Figure~\ref{appfig:Figure_7}(a), the time origin is chosen at the moment when the MW drive is switched off. The interval before this point, labeled $T_{\mathrm{dc}}$, therefore corresponds to the MW-on time for the respective duty cycle.

The recovery part of each trace is fit with a single exponential,
\begin{equation}
    I(t)=I_0 + A\left[\exp\left(\frac{t-t_0}{\tau_{\mathrm{th}}}\right)\right],
    \label{appeq:thermal_exp_fit}
\end{equation}
where $\tau_{\mathrm{th}}$ is an effective thermal relaxation time. Only the rising part of the fluorescence recovery, indicated by the solid lines in Fig.~\ref{appfig:Figure_7}(a), is included in the fit. A full quantitative treatment would require solving the heat equation for the complete sample stack, including the diamond membrane, niobium film, adhesive layer, bonding wires, and sample holder.
This is not attempted here because the relevant thermal parameters are strongly temperature dependent at cryogenic temperatures and are not precisely known for the UV-cured adhesive layer. In addition, the heating pulse drives the system far from the small-temperature-excursion limit, since the local diamond temperature can exceed the calibrated optical range during the MW-on interval.
The exponential fit should therefore be understood as a phenomenological description of the observed recovery rather than as a microscopic heat-transport model.

Despite this simplification, the exponential model captures the measured fluorescence revival well and yields a characteristic relaxation time on the order of $0.5\,\si{\milli\second}$. Figure~\ref{appfig:Figure_7}(b) summarizes the extracted relaxation times as a function of duty cycle and gives an effective time constant of
\begin{equation}
    \tau_{\mathrm{th}}=\SI{0.54(17)}{\milli\second}.
\end{equation}

This millisecond-scale relaxation is much slower than the internal thermal equilibration expected within the diamond membrane itself. It reflects the thermal response of the surrounding sample assembly, in particular the adhesive layer between the diamond and the chip carrier.
Compared to diamond, polymeric adhesives have a much larger heat capacity and much lower thermal conductivity at cryogenic temperatures, allowing them to store heat after superconducting breakdown and release it only slowly to the sample holder.

Table~\ref{tab:temperature-materials} summarizes representative low-temperature material parameters relevant to this thermal picture. The values are order-of-magnitude estimates rather than precise parameters of the present device, since they depend strongly on purity, microstructure, thickness, curing conditions, and interfaces.
The main point is the large contrast between diamond and the surrounding materials: heat can spread rapidly through the diamond membrane, but the overall thermalization is limited by the lower-diffusivity, higher-heat-capacity interface materials.

\begin{table*}
\centering
\caption{\label{tab:temperature-materials}
Representative low-temperature material parameters for the interface materials considered in this work. The table summarizes order-of-magnitude values of the specific heat $c_p$, thermal conductivity $\kappa$, and thermal diffusivity $\alpha$ for (single crystal) diamond, epoxy/resin, Kapton, and gold at $T=0.1$, $1$, and $10\,\si{\kelvin}$, enabling a direct comparison of their thermal properties across the relevant temperature range. Sources are summarized in the notes below. Abbreviations: T. = Temperature}
\begin{tabular}{p{1.8cm} p{2.3cm} p{2.6cm} p{2.3cm} p{2.2cm}}
    \hline
    \hline
    & \multicolumn{4}{c}{\textbf{Material}} \\
    \cline{2-5}
    \textbf{T. (K)} & \textbf{Diamond} & \textbf{Epoxy/Resin} & \textbf{Kapton} & \textbf{Gold} \\
    \hline
    \hline
    & \multicolumn{4}{c}{Specific heat $c_p$ (\si{\joule\per\kilogram\per\kelvin})} \\
    \cline{2-5}
    \hline
    $0.1$ & $10\text{--}20 \times 10^{-9}$ & $20\text{--}40 \times 10^{-6}$ & - & $40 \times 10^{-6}$ \\
    
    $1$ & $10\text{--}20 \times 10^{-6}$ & $20\text{--}40 \times 10^{-3}$ & - & $2.9 \times 10^{-3}$ \\
    
    $10$ & $10\text{--}30 \times 10^{-3}$ & $10\text{--}20$ & 11.68 & 2.18 \\
    \hline
    & \multicolumn{4}{c}{Thermal conductivity $\kappa$ (\si{\watt\per\meter\per\kelvin})} \\
    \cline{2-5}
    \hline
    $0.1$ & $1.5\text{--}3.5 \times 10^{-3}$ & $0.2\text{--}0.5 \times 10^{-3}$ & $0.13\text{--}0.55\times 10^{-3}$ & $2\text{--}40$ \\
    
    $1$ & $0.5\text{--}2$ & $20\text{--}40 \times 10^{-3}$ & $1.2\text{--}5.2\times 10^{-3}$ & $25\text{--}350$ \\
    
    $10$ & $300\text{--}1000$ & $30\text{--}60 \times 10^{-3}$ & $12\text{--}50\times 10^{-3}$ & $300\text{--}2800$ \\
    \hline
    & \multicolumn{4}{c}{Thermal diffusivity $\alpha$ (\si{\meter\squared\per\second})} \\
    \cline{2-5}
    \hline
    $1$ & $7\text{--}50$ & $0.5\text{--}1 \times 10^{-3}$ & - & $0.4\text{--}6.2$ \\
    $10$ & $4\text{--}15$ & $1.6\text{--}3.3 \times 10^{-6}$ & $0.7\text{--}3 \times 10^{-6}$ & $7\text{--}66 \times 10^{-3}$ \\
    \hline
    \hline
\end{tabular}
\vspace{1mm}
\begin{minipage}{0.95\textwidth}
\footnotesize
\textit{Notes:} Values are order-of-magnitude estimates compiled from literature sources. 
Diamond data from Refs.~\cite{Vasiliev_2010, Inyushkin_2018, Desnoyehs_1958, Reeber_1996, DeSorbo_1953};
epoxy data from Refs.~\cite{Anderson_1963, Araujo_1976, Kelham_Rosenberg_1981}; 
resin data from Refs.~\cite{Nakamura_2018, Freeman_1986}; 
Kapton data from Ref.~\cite{Bradley_Radebaugh_2013, NIST_Kapton, Benford_1999, Lawrence_2000, Radebaugh_1973}; 
gold data from Refs.~\cite{White_1953, Martin_1966, Geballe_1952, Rosenberg_1955}.
Note, that the thermal conductivity of diamond and gold are highly dependent on purity and thickness. Diamond values refer to synthetic single-crystalline diamond. For diamond films we refer to \cite{Morelli_1988, Anthony_1991}.
\end{minipage}
\end{table*}